\documentclass[preprint,aps,superscriptaddress,showpacs,floatfix]{revtex4}
\usepackage{graphicx}
\begin{document}
\title{Cross sections for pentaquark baryon production from protons in
reactions induced by hadrons and photons}
\bigskip
\author{W. Liu}
\affiliation{Cyclotron Institute and Physics Department, Texas A\&M
University, College Station, Texas 77843-3366, USA}
\author{C. M. Ko}
\affiliation{Cyclotron Institute and Physics Department, Texas A\&M
University, College Station, Texas 77843-3366, USA}

\date{\today}

\begin{abstract}
Using hadronic Lagrangians that include the interaction of
pentaquark $\Theta^+$ baryon with $K$ and $N$, we evaluate the
cross sections for its production from meson-proton,
proton-proton, and photon-proton reactions near threshold. With
empirical coupling constants and form factors, the predicted cross
sections are about 1.5 mb in kaon-proton reactions, 0.1 mb in
rho-nucleon reactions, 0.05 mb in pion-nucleon reactions, 20
$\mu$b in proton-proton reactions, and 40 nb in photon-proton
reactions.
\end{abstract}


\maketitle

\section{introduction}
Recently, a narrow baryon state was inferred from  the invariant mass
spectrum of $K^+n$ or $K^0p$ in nuclear reactions induced by photons
\cite{nakano,stepanyan} or kaons \cite{barmin}. The extracted mass of
about 1.54 GeV and width of less than 21-25 MeV are consistent with those
of the pentaquark baryon $\Theta^+$ consisting of $uudd\bar s$ quarks
predicted in the chiral soliton model \cite{diakonov}. Its existence
has also been verified recently in the constituent quark model
\cite{riska,lipkin} and the QCD sum rules \cite{zhu}. Although the
spin and isospin of $\Theta^+$ are predicted to be 1/2 and 0,
respectively, those of the one detected in experiments are not yet determined.
Studies have therefore been carried out to predict its decay branching
ratios based on different assignments of its spin and isospin \cite{carl,ma}.
Since both kaons and nucleons are present in the hadronic
matter formed in relativistic heavy ion collisions, the number of
$\Theta^+$ produced in these collisions may be appreciable. Using the
statistical model, which assumes that $\Theta^+$ baryons
are in chemical equilibrium with other hadrons, Randrup
\cite{randrup} has estimated its abundance and finds that the
$\Theta/\Lambda$ ratio in the midrapidity is about 10-20\% in central
Au+Au collisions at $\sqrt{s_{NN}}=200$ GeV available from the
Relativistic Heavy Ion Collider (RHIC). With about 7 $\Lambda$'s produced
in midrapidity \cite{star,phenix}, one expects that there would be about
one midrapidity $\Theta^+$ present in these collisions. Since a quark-gluon
plasma is believed to have formed in the initial stage of relativistic
heavy ion collisions, $\Theta^+$ baryons can also be produced during
the hadronization of the quark-gluon plasma. This contribution has
been studied in Ref.\cite{chen} using the quark coalescence model and
was found to be important as production from later hadronic matter
is less significant due to the small coupling between $\Theta^+$ and other
hadrons as predicted in the chiral soliton model \cite{diakonov}.
The coupling may even be smaller as reanalysis of $K^+p$ and $K^+d$
elastic scattering data has indicated that they are consistent with
the existence of resonances with width of only 1 MeV \cite{arndt}.
In this case, $\Theta^+$ baryon, like other multistrange baryons
\cite{rafelski}, can also be used as a signal for the quark-gluon plasma
in relativistic heavy ion collisions. On the other hand, if the
$\Theta^+$ baryon interacts strongly in hadronic matter, its final
number in relativistic heavy ion collisions would become independent of its
initial number produced from the quark-gluon plasma. Therefore,
knowledge of the cross sections for $\Theta^+$ production and
absorption by hadrons is important for understanding the mechanism for
its production in relativistic heavy ion collisions.

In this paper, we evaluate the cross sections for $\Theta^+$ production
from nucleons induced by mesons and protons using a hadronic model
that is based on $SU(3)$ flavor symmetry with empirical hadron masses
and form factors at interaction vertices. For the coupling constant
between $\Theta^+$ and $KN$, it is determined from the width of $\Theta^+$.
Introducing the photon as a $U_{\rm em}(1)$ gauge boson, we extend the
hadronic model to also calculate the cross section for $\Theta^+$
production from photon-proton reactions, which is relevant to the experiments
in which $\Theta^+$ was detected \cite{nakano,stepanyan}.

This paper is organized as follows. In Sect.\ref{mesonnucleon},
the cross sections for $\Theta^+$ production from meson-nucleon
reactions are evaluated. Production of $\Theta^+$ from proton-proton
reactions are studied in Sect. \ref{protonproton} with inclusion of
both two-body and three-body final states. The cross section for
$\Theta^+$ production from photon-proton reactions is then determined
in Sect. \ref{photon}. Finally, a brief summary is given in Sect.
\ref{summary}.

\section{theta production from meson-nucleon reactions}
\label{mesonnucleon}

\begin{figure}[ht]
\includegraphics[height=3.5in,width=3.5in,angle=0]{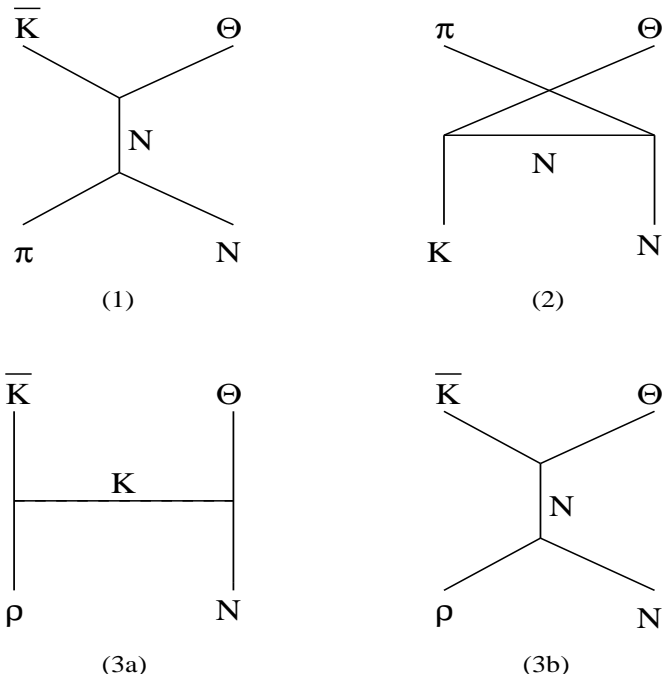}
\caption{Diagrams for $\Theta^+$ production from meson-nucleon reactions.}
\label{diagram1}
\end{figure}

Possible reactions for $\Theta^+$ production in meson-nucleon interactions
are $\pi N\to\bar K\Theta$, $KN\to\pi\Theta$, and $\rho N\to\bar K\Theta$
as shown by diagrams in Fig.\ref{diagram1}. The interaction Lagrangians
needed to evaluate the cross sections for these reactions are:
\begin{eqnarray}
{\cal L}_{KN\Theta}&=&ig_{KN\Theta}(\bar\Theta\gamma_5N\bar K
+\bar N\gamma_5 \Theta K),\nonumber\\
{\cal L}_{\pi NN}&=&-ig_{\pi NN}\bar N\gamma_5\pi N,\nonumber\\
{\cal L}_{\rho NN}&=& g_{\rho NN}\bar
N\left(\gamma^\mu\rho+\frac{\kappa_\rho}{2m_N}\sigma^{\mu\nu}
\partial_\mu\rho_\nu\right)N,\nonumber\\
{\cal L}_{\rho KK}&=&ig_{\rho KK}(K\rho^\mu\partial_\mu\bar
K-\partial_\mu K\rho^\mu\bar K).
\label{lagrangian}
\end{eqnarray}
In the above, $\sigma^{\mu\nu}=i\{\gamma^\mu,\gamma^\nu\}/2$ with
$\gamma^\mu$ denoting the Dirac gamma matrices; the isospin doublet
kaon and nucleon field are denoted by $N$ and $K$, respectively;
while the isospin triplet pion and rho meson fields are given by
$\pi=\vec\tau\cdot\vec\pi$ and $\rho^\mu=\vec\tau\cdot\vec\rho^\mu$,
respectively, with $\vec\tau$ denoting the Pauli spin matrices.
The spin 1/2 and isospin 0 pentaquark baryon is denoted by the $\Theta$ field.
For coupling constants involving normal hadrons, they are taken to be
$g_{\pi NN}=13.5$, $g_{\rho NN}=3.25$, $g_{\rho KK}=3.25$,
$\kappa_\rho=6.1$ as usually used in hadronic models \cite{li}.
The coupling constant $g_{KN\Theta}$ between $\Theta^+$ and $NK$ is
determined from its width given by
\begin{equation}
\Gamma_\Theta=\frac{g_{KN\Theta}k}{2\pi}\frac{\sqrt{m_N^2+k^2}-m_N}
{m_\Theta},
\end{equation}
where $m_N$ and $m_\Theta$ are the masses of nucleon and $\Theta^+$,
respectively, and $k$ is the center-of-mass momentum of $N$ and $K$
in the rest frame of $\Theta^+$. Using $m_\Theta=1.54$ GeV and
$\Gamma_\Theta=20$ MeV, we find $g_{KN\Theta}=4.4$, which is
comparable to that given by the chiral soliton model \cite{diakonov}.

The amplitudes for the three reactions shown in Fig.\ref{diagram1} are
\begin{eqnarray}
{\cal M}_1&=&g_{\pi
NN}g_{KN\Theta}\bar\Theta(p_4)\frac{m_N-{p\mkern-8mu/}_1-{p\mkern-8mu/}_2}
{s-m^2_N}N(p_2),\nonumber\\
{\cal M}_2&=&g_{\pi NN}g_{KN\Theta}\bar\Theta(p_4)\frac{m_N-{p\mkern-8mu/}_2
+{p\mkern-8mu/}_3}{u-m^2_N}N(p_2),\nonumber\\
{\cal M}_{3a}&=&ig_{\rho KK}g_{KN\Theta}\bar\Theta(p_4)\gamma_5
N(p_2)\frac{1}{t-m^2_K}(2p_3-2p_1)^\mu\epsilon_\mu,\nonumber\\
{\cal M}_{3b}&=&ig_{\rho NN}g_{KN\Theta}\bar\Theta(p_4)\gamma_5
\frac{{p\mkern-8mu/}_1+{p\mkern-8mu/}_2+m_N}{s-m^2_N}
\left(\gamma^\mu-i\frac{\kappa_\rho}{2m_N}\sigma^{\nu\mu}p_{1\nu}\right)
N(p_2)\epsilon_\mu,
\label{amplitude}
\end{eqnarray}
where $\epsilon_\mu$ is the polarization vector of $\rho$ meson.
In the above, $p_1$ and $p_2$ denote the momenta of initial state
particles while $p_3$ and $p_4$ denote those of final state particles
on the left and right side of a diagram. The usual Mandelstam
variables are given by $s=(p_1+p_2)^2$, $t=(p_1-p_3)^2$, and $u=(p_1-p_4)^2$.

The spin and isospin averaged
cross sections for these reactions can be written as
\begin{eqnarray}
\frac{d\sigma}{dt}=\frac{1}{64\pi sp^2_i}\frac{1}{S_i}\frac{1}{I_i}
\sum_{s,I}|{\cal M}_i|^2.
\label{sigma}
\end{eqnarray}
In the above, $S_i$ and $I_i$ denote spin and isospin factors,
and their values are 2 and 6 for the reaction $\pi N\to\bar K\Theta$,
2 and 4 for the reaction $KN\to\pi\Theta$, and 6 and 6 for the
reaction $\rho N\to\bar K\Theta$.

To evaluate the cross sections for these reactions, form factors are
needed at interaction vertices to take into account the finite sizes of
hadrons. We adopt the monopole form factor used in
Refs.\cite{lin,di,liu1,liu2}, i.e.,
\begin{eqnarray}
F({\bf q}^2)=\frac{\Lambda^2}{\Lambda^2+{\bf q}^2}
\label{form},
\end{eqnarray}
with ${\bf q}^2$ denoting either the squared three momentum of external
particles in $s$ and $u$ channel diagrams or the squared three
momentum transfer in $t$ channel diagrams. The cutoff parameter
is taken to be $\Lambda=0.5$ GeV for all interaction vertices, based on
fitting the measured cross section for the reaction $\pi N\to K\Lambda$
using similar hadronic Lagrangians \cite{liu3}.

\begin{figure}[ht]
\includegraphics[width=3.5in,height=4.5in,angle=-90]{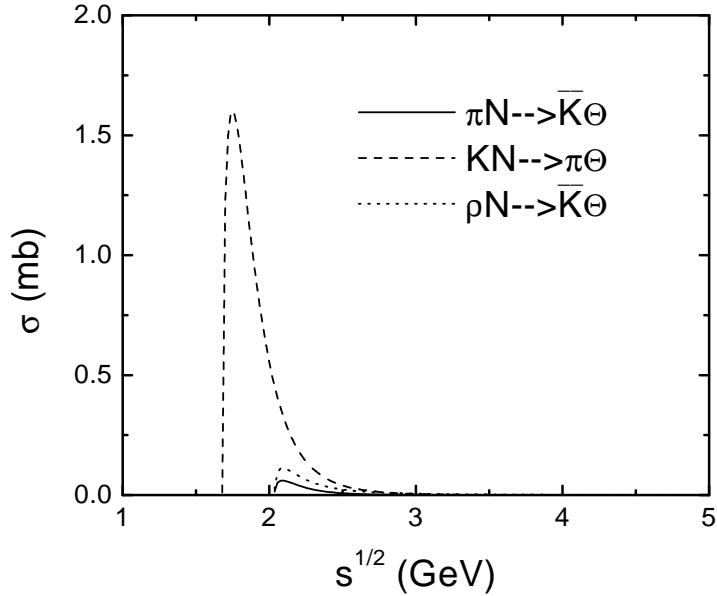}
\caption{Spin and isospin averaged cross sections for $\Theta^+$ production
from meson-nucleon reactions as functions of center-of-mass energy:
$\pi N\to\bar K\Theta$ (solid curve), $KN\to\pi\Theta$ (dashed curve),
and $\rho N\to\bar K\Theta$ (dotted curve).}
\label{crossds1}
\end{figure}

The resulting spin and isospin averaged cross sections are shown in
Fig.\ref{crossds1} as functions of center-of-mass energy. It is seen
that all cross sections peak at energies slightly above their
threshold. The largest cross section is from the reaction
$KN\to\pi\Theta$, with peak value of about 1.6 mb as result of
small mass difference between initial and final states. The cross
sections for other two reactions $\pi N\to \bar K\Theta$ and
$\rho N\to\bar K\Theta$ are much smaller, with values about 0.05 mb
and 0.1 mb, respectively.

\section{theta production from proton-proton reactions}
\label{protonproton}

\begin{figure}[ht]
\includegraphics[width=4.0in,height=2.5in,angle=0]{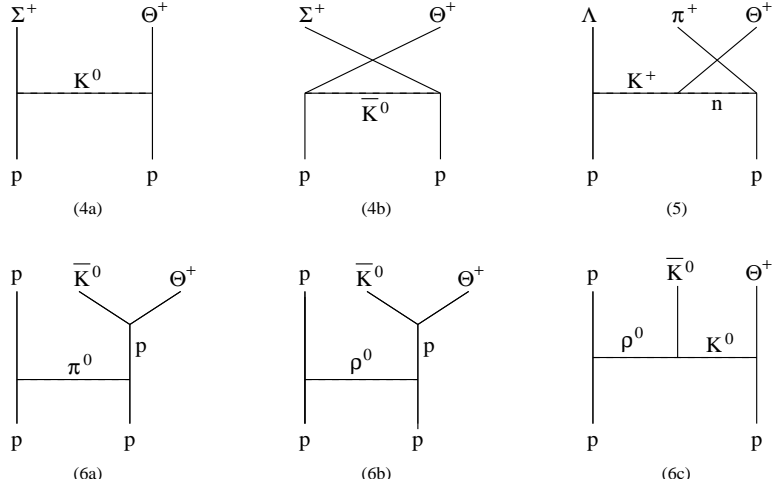}
\caption{Diagrams for $\Theta$ production from proton-proton reactions.}
\label{diagram2}
\end{figure}

The $\Theta^+$ baryon can also be produced from proton-proton reactions
$pp\to\Sigma^+\Theta^+$ with two-body final state as well as
$pp\to\pi^+\Lambda\Theta^+$ and $pp\to\bar K^0 p\Theta^+$
with three-body final states.  These reactions are shown by the diagrams
in Fig.\ref{diagram2}. Additional interaction Lagrangians besides
those given in Eq.(\ref{lagrangian}) needed to evaluate their cross
sections are
\begin{eqnarray}
{\cal L}_{KN\Lambda}&=&ig_{KN\Lambda}(\bar K\bar \Lambda\gamma_5
N+\bar N\gamma_5\Lambda\bar K),\nonumber\\
{\cal L}_{KN\Sigma}&=&ig_{KN\Sigma}\bar
N\gamma_5\bar{\Sigma}K+ {\rm H.c.},
\end{eqnarray}
where $\Lambda$ and $\Sigma=\vec\tau\cdot\vec\Sigma$ denote the
isospin singlet and triplet hyperons, respectively. The coupling constants
$g_{KN\Lambda}$ and$g_{KN\Sigma}$ are related to $g_{\pi NN}$
via the $SU(3)$ relations \cite{liu1,li}:
\begin{eqnarray}
g_{KN\Lambda}&=&\frac{3-2\alpha_D}{\sqrt{3}}g_{\pi NN}\simeq 13.5,\nonumber\\
g_{KN\Sigma}&\simeq&(1-2\alpha_D)g_{\pi NN}=-3.78,
\end{eqnarray}
with $\alpha_D=D/(D+F)=0.64$.

For the reaction $pp\to\Sigma^+\Theta^+$ with two-body final state,
the two amplitudes are
\begin{eqnarray}
{\cal M}_{4a}&=&-g_{KN\Theta}g_{KN\Sigma}\bar \Sigma(p_3)\gamma_5
p(p_1))\frac{1}{t-m^2_K}\bar\Theta(p_4)\gamma_5 p(p_2),\nonumber\\
{\cal M}_{4b}&=&-g_{KN\Theta}g_{KN\Sigma}\bar \Sigma(p_3)\gamma_5
p(p_2)\frac{1}{u-m^2_K}\bar\Theta(p_4)\gamma_5 p(p_1).
\end{eqnarray}
Its cross section can be evaluated using Eq.(\ref{sigma}) without
isospin factor.

For the reactions $pp\to\pi^+\Lambda\Theta^+$ and $pp\to\bar K^0p\Theta^+$
with three-body final states, their amplitudes can be written in terms
of those of off-shell two-body subprocesses involving the
exchanged meson and the proton on the right side of a diagram
\cite{liu1,liu2}, i.e.,
\begin{eqnarray}
{\cal M}_5 & = &
ig_{KN\Lambda}\bar{\Lambda}(p_{3})\gamma_{5}p(p_{1})
\frac{1}{t-m^{2}_{K}}{\cal M}_{K^+p\to\pi^+\Theta^+},\\
{\cal M}_6 & = & -ig_{\pi NN}\bar{p}(p_{3})\gamma_{5}p(p_{1})
\frac{1}{t-m^{2}_{\pi}}{\cal M}_{\pi^0p\to\bar K^0\Theta^+}
\nonumber\\
&&+ g_{\rho NN}\bar{p}(p_{3})\left[\gamma^{\mu}
+i\frac{\kappa_{\rho}}{2m_{N}}\sigma^{\alpha\mu}
(p_{1}-p_{3})_{\alpha}\right]p(p_1)\nonumber\\
&&\times \left[-g_{\mu\nu}+\frac{(p_1-p_3)_{\mu}
(p_1-p_3)_{\nu}}{m^{2}_{\rho}}\right]
\frac{1}{t-m^{2}_{\rho}}{\cal M}_{\rho^0p\to\bar K^0\Theta^+},
\end{eqnarray}
where ${\cal M}_{K^+p\to\pi^+\Theta^+}$,
${\cal M}_{\pi^0p\to\bar K^0\Theta^+}$, and
${\cal M}_{\rho^0p\to\bar K^0\Theta^+}$ are the amplitudes for
the subprocesses $K^+p\to\pi^+\Theta^+$, $\pi^0p\to\bar K^0\Theta^+$,
and $\rho^0p\to\bar K^0\Theta^+$. In terms of their off-shell cross
sections, the cross sections for $\Theta^+$ production in proton-proton
reactions with three-body final states are given by
\begin{eqnarray}
\frac{d\sigma_{pp\to\pi^+\Lambda\Theta^+}}{dtds_1}&=&\frac{g^{2}_{KN\Lambda}}
{32\pi^{2}sp^{2}_{i}}k\sqrt{s_{1}}
[-t+(m_N-m_{\Lambda})^2]\frac{F({\bf q}^2)}{(t-m^{2}_{K})^{2}}
\sigma_{K^+p\to \pi^+\Theta^+}(s_{1},t),\\
\frac{d\sigma_{pp\to\bar K^0 p\Theta^+}}{dtds_1}&=&\frac{g^{2}_{\pi NN}}
{32\pi^{2}sp^{2}_{i}}k\sqrt{s_{1}}(-t)\frac{f(t)}{(t-m^{2}_{\pi})^{2}}
\sigma_{\pi^0 p\to \bar{K}^0\Theta^+}(s_{1},t),\nonumber\\
&&+\frac{3g^{2}_{\rho NN}}{64\pi^{2}sp^{2}_{i}}k\sqrt{s_{1}}
\frac{f(t)}{(t-m^{2}_{\rho})^{2}}\left[4(1+\kappa_{\rho})^2(-t-2m^{2}_{N})-
\kappa^{2}_{\rho}\frac{(4m^{2}_{N}-t)^{2}}{2m^{2}_{N}}\right.\nonumber\\
&&\left.+4(1+\kappa_{\rho})\kappa_{\rho}(4m^{2}_{N}-t)\right]
\sigma_{\rho^0p\to\bar{K}^0\Theta^+}(s_{1},t). \label{threebody}
\end{eqnarray}
In the above, $p_i$ is the center-of-mass momentum of two initial protons,
$t$ is the squared four momentum transfer of exchanged meson, $s$
is the squared total center-of-mass energy, and $s_{1}$ and $k$ are,
respectively, the squared invariant mass and center-of-mass
momentum of exchanged meson in the subprocess. We note that there is no
interference between amplitudes involving exchange of pion and rho meson.

For form factors at $\pi NN$ and $\rho NN$ vertices involving pion meson
or rho meson exchange, we use the usual covariant monopole form factor
\cite{liu1}
\begin{eqnarray}
f(t)=\frac{\Lambda^2-m^2_{\rm ex}}{\Lambda^2-t}
\end{eqnarray}
with $m_{\rm ex}$ denoting the mass of exchanged meson. The cutoff
parameter $\Lambda$ is taken to have value 1.3 GeV for pion exchange
and 1.4 GeV for rho exchange as in Ref.\cite{liu1}. At other interaction
vertices, we use monopole form factors $F({\bf q}^2)$ of Eq.(\ref{form})
with same cutoff parameter $\Lambda=0.5$ GeV as introduced in
Sect.\ref{mesonnucleon}.

\begin{figure}[ht]
\includegraphics[width=3.5in,height=4.5in,angle=-90]{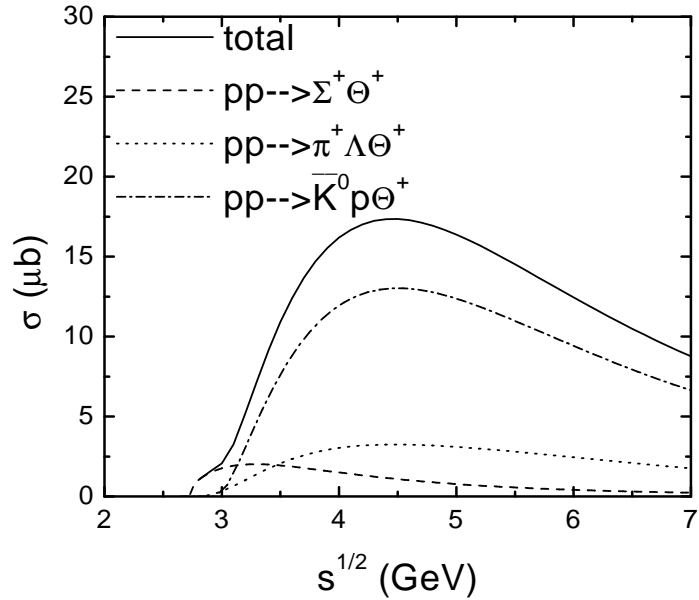}
\caption{Cross sections for $\Theta^+$ production
from proton-proton reactions $pp\to\Sigma^+\Theta^+$ (dashed curve),
$pp\to\pi^+\Lambda\Theta^+$ (dotted curve), and $pp\to\bar K^0 p\Theta^+$
(dash-dotted curve) functions of center-of-mass energy. The total
cross section is given by solid curve.}
\label{cross2}
\end{figure}

In Fig.\ref{cross2}, we show the cross sections for $\Theta^+$ production
from proton-proton reactions with either two-body or three-body
final states as functions of center-of-mass energy. It is seen that the
cross section for the reaction $pp\to\Sigma^+\Theta^+$ (dashed curve) with
two-body final state has a peak value about 2 $\mu b$ around 3.3 GeV.
This cross section is comparable to that for the reaction
$pp\to\pi^+\Lambda\Theta^+$ (dotted curve) with three-body final
state, which is about 3 $\mu$b at center-of-mass energy 4.2 GeV.
These two cross sections are much smaller than the peak cross section
of 13 $\mu$b for the reaction $pp\to\bar K^0p\Theta^+$ (dash-doted
curve) with three-body final state. The total cross section
for $\Theta^+$ production from proton-proton reactions obtained from
the sum of above three partial cross sections is shown
by the solid curve, and its value is almost two order-of-magnitude
larger than that estimated in Ref.\cite{cassing} based on the phase-space
argument.

\section{theta production from photon-proton reactions}
\label{photon}

\begin{figure}[ht]
\includegraphics[width=3.0in,height=1.5in,angle=0]{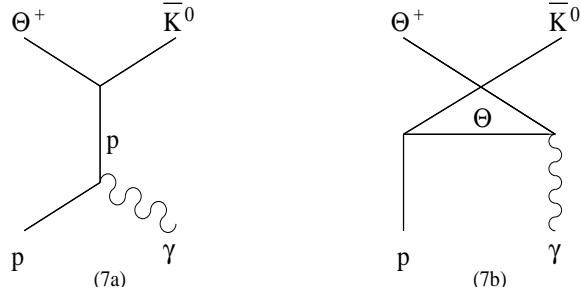}
\caption{Diagrams for $\Theta^+$ production from photon-proton reactions
with two-body final state.}
\label{diagram3}
\end{figure}

\begin{figure}[ht]
\includegraphics[width=4.5in,height=3.0in,angle=0]{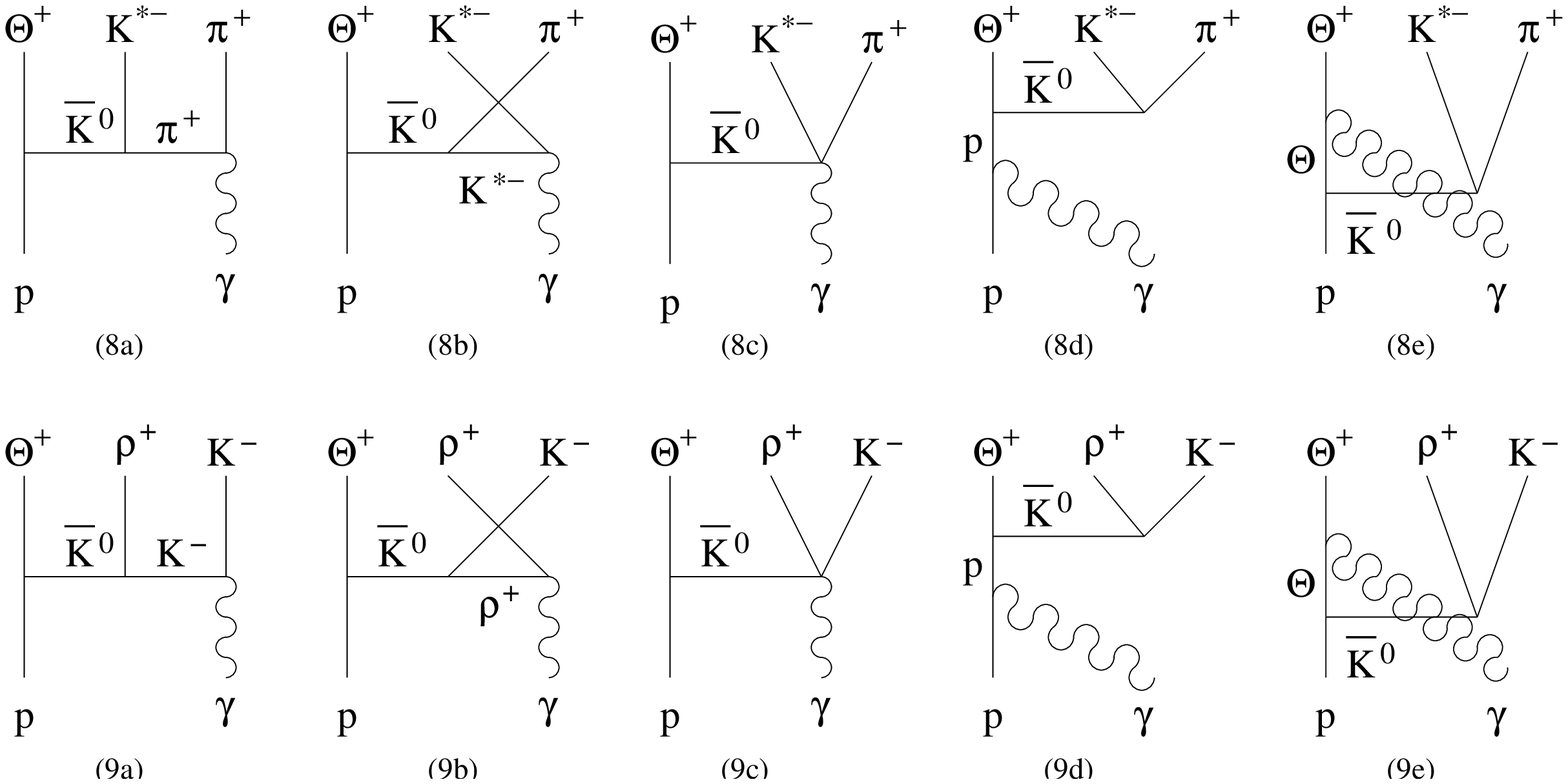}
\caption{Diagrams for $\Theta^+$ production from photon-proton reactions
with three-body final states.}
\label{diagram4}
\end{figure}

The hadronic Lagrangians introduced in previous sections can be
generalized to include photons in order to evaluate the cross section
for $\Theta^+$ production from photon-proton reactions. This
includes the reaction $\gamma p\to\bar K^0\Theta^+$ with
two-body final state as shown by the diagrams in Fig.\ref{diagram3}
as well as the reactions $\gamma p\to K^{*-}\pi^+\Theta^+$ and
$\gamma p\to K^-\rho^+\Theta^+$ with three-body final states as
shown by the diagrams in Fig.\ref{diagram4}.

To evaluate the cross sections for these reactions, we need the following
interaction Lagrangians involving photons:
\begin{eqnarray}
{\cal L}_{\gamma NN}&=&ieA^\mu\bar N\gamma_\mu[(1+\tau_3)/2]N,\nonumber\\
{\cal L}_{\gamma \Theta\Theta}&=&ieA^\mu\bar \Theta\gamma_\mu\Theta,\nonumber\\
{\cal L}_{\gamma\pi\pi}&=&eA^\mu(\partial_\mu\vec\pi\times\vec\pi)_3,
\nonumber\\
{\cal L}_{\gamma \rho\rho} & = & e\{A^{\mu}
(\partial_{\mu}\vec\rho^{\nu}\times\vec{\rho}_{\nu})_3
+[(\partial_{\mu}A^\nu \vec\rho_{\nu}-A^{\nu}
\partial_{\mu}\vec\rho_{\nu})\times\vec\rho^\mu]_3+[\vec\rho^\mu\times(A^{\nu}
\partial_{\mu}\vec\rho_\nu-\partial_{\mu}A^{\nu}\vec\rho_{\nu})]_3\},
\nonumber\\
{\cal L}_{\gamma KK}&=&ieA^\mu[KQ\partial_\mu\bar K-\partial_\mu KQ\bar K],
\end{eqnarray}
where $A_\mu$ denotes the photon field and $Q$ is the diagonal charge
operator with elements 0 and -1. Also needed is the interaction
Lagrangian between $\pi$ and $KK^*$, which is given by
\begin{equation}
{\cal L}_{\pi KK^*}=ig_{\pi KK^*}K^{*\mu}(\bar K\partial_\mu\pi
-\partial_\mu\bar K\pi)+{\rm H.c.},
\end{equation}
with the coupling constant $g_{\pi KK^*}=3.28$ determined from the decay
width of $K^*$.

For the reaction $\gamma p\to\bar K\Theta^+$ with two-body final state,
its amplitude is
\begin{eqnarray}
{\cal M}_7={\cal M}_{7a}+{\cal M}_{7b}
\end{eqnarray}
with
\begin{eqnarray}
{\cal M}_{7a} & = &-eg_{KN\Theta}\frac{1}{s-m^2_N}
\bar{\Theta}(p_3)\gamma^5({p\mkern-10mu/}_1+{p\mkern-10mu/}_2+m_N)
\gamma^\mu N(p_1)\epsilon_\mu,\nonumber\\
{\cal M}_{7b} & = &-eg_{KN\Theta}\frac{1}{u-m^{2}_{\Theta}}
\bar{\Theta}(p_3)\gamma^\mu({p\mkern-10mu/}_1-{p\mkern-10mu/}_4
+m_{\Lambda_c})\gamma^5 N(p_1)\epsilon_\mu,
\end{eqnarray}
where $\epsilon_\mu$ is the polarization vector of $\gamma$.

For form factors, an overall one is multiplied to the total amplitude
of each reaction in order to maintain gauge invariance for the
resulting amplitude, as in Ref.\cite{liu2} for photoproduction of
charmed hadrons on protons. This form factor is taken to have the
monopole form of Eq.(\ref{form}) but with ${\bf q}$ denoting the
photon three momentum in center-of-mass system. As in photoproduction of
charmed hadrons, we use a cutoff parameter $\Lambda=0.75$ GeV.
The cross section for the reaction $\gamma p\to\Sigma^+\Theta^+$
with two-body final state can be similarly written as that of Eq.(\ref{sigma})
without isospin factor.

For the reactions $\gamma p\to K^-\pi^+\Theta^+$ and
$\gamma p\to K^-\rho^+\Theta^+$ with three-body final states via
$\bar K^0$ exchange, the photon couples either to $K$ meson as in
first three diagrams or to external baryons as in other two
diagrams. Since the contributions from latter diagrams are much
smaller than those from the former diagrams, as shown explicitly in
charmed hadron production from proton-proton reactions with three-body
final states \cite{liu2}, they are thus neglected in following
calculations. As a result, results obtained in present study for
$\Theta^+$ production with three-body final states violate
slightly the gauge invariance. The amplitudes for the two reactions
with three-body final states can then be written as
\begin{eqnarray}
{\cal
M}_i&=&ig_{KN\Theta}\bar\Theta(p_3)\gamma_5N(p_1)\frac{1}{t-m^2_K}
({\cal M}^{sub}_{ia}+{\cal M}^{sub}_{ib}+{\cal M}^{sub}_{ic}),
\end{eqnarray}
with $i=8$ and 9 denote, respectively, the reaction
$\gamma p\to K^-\pi^+\Theta^+$ and the reaction
$\gamma p\to K^-\rho^+\Theta^+$ in Fig.\ref{diagram4}.
In the above, the amplitudes ${\cal M}^{sub}_{ia}$, ${\cal M}^{sub}_{ib}$,
and ${\cal M}^{sub}_{ic}$ are for the subprocesses
$\gamma \bar K^0\to \pi^+ K^{*-}$ and $\gamma \bar K^0\to \rho^+ K^-$,
and they are given explicitly by
\begin{eqnarray}
{\cal M}^{sub}_{8a}&=&\sqrt{2}eg_{\pi KK^*}(-2k_1 +k_3)^\mu
\frac{1}{(k_1-k_3)^2-m^2_K}\times(k_1 -k_3 +k_4)^\nu\varepsilon_{3\mu}
\varepsilon_{2\nu},\nonumber\\
{\cal M}^{sub}_{8b}&=&-\sqrt{2}eg_{\pi KK^*}(-k_1-k_4)^\alpha
\frac{1}{(k_1-k_4)^2-m^2_{K^*}}\left[g_{\alpha\beta}
-\frac{(k_1-k_4)_\alpha(k_1-k_4)_\beta}{m^2_{K^*}}\right]\nonumber\\
&&\times[(-k_2 -k_3)^\beta g^{\mu\nu}+(-k_1+k_2 +k_4)^\nu g^{\beta\mu}
+(k_1+k_3-k_4)^\mu g^{\beta\nu}]\varepsilon_{3\mu}\varepsilon_{2\nu},
\nonumber\\
{\cal M}^{sub}_{8c} &=&\sqrt{2}eg_{\pi KK^*}g^{\mu\nu}\varepsilon_{3\mu}
\varepsilon_{2\nu},\nonumber\\
{\cal M}^{sub}_{9a}&=&\sqrt{2}eg_{\rho KK}(-2k_1 +k_3)^\mu\frac{1}
{(k_1-k_3)^2-m^2_K}(k_1 -k_3 +k_4)^\nu\varepsilon_{3\mu}
\varepsilon_{2\nu},\nonumber\\
{\cal M}^{sub}_{9b} &=&-\sqrt{2}eg_{\rho KK}(-k_1-k_4)^\alpha
\frac{1}{(k_1-k_4)^2-m^2_{\rho}}\left[g_{\alpha\beta}
-\frac{(k_1-k_4)_\alpha(k_1-k_4)_\beta}{m^2_{\rho}}\right]\nonumber\\
&&\times[(-k_2 -k_3)^\beta g^{\mu\nu}+(-k_1+k_2 +k_4)^\nu
g^{\beta\mu}+(k_1+k_3-k_4)^\mu g^{\beta\nu}]\varepsilon_{3\mu}
\varepsilon_{2\nu},\nonumber\\
{\cal M}^{sub}_{9c}&=&\sqrt{2}eg_{\rho
KK}g^{\mu\nu}\varepsilon_{3\mu} \varepsilon_{2\nu}.
\end{eqnarray}
In the above, $k_1$ and $k_4$ are the momenta of initial
$K$ and final pseudoscalar meson, while $k_2$ and $k_3$ are
those of initial photon and final vector meson with their polarization
vectors denoted by $\epsilon_2$ and $\epsilon_3$, respectively.

As in proton-proton reactions, cross sections for the two reactions
$\gamma p\to K^{*-}\pi^+\Theta^+$ and $\gamma p\to K^-\rho^+\Theta^+$
with three-body final states can be expressed in terms of off-shell cross
sections for the subprocesses $\gamma \bar K^0\to K^{*-}\pi^+$
and $\gamma\bar K^0\to\rho^+K^-$ involving two particles in final
states, i.e.,
\begin{eqnarray}
\frac{d\sigma_{\gamma p\to K^{*-}\pi^+\Theta^+}}{dtds_1}
&=&\frac{g^{2}_{KN\Theta}}{32\pi^{2}sp^{2}_{i}}k\sqrt{s_{1}}
[-t+(m_N-m_{\Theta})^2]\frac{F({\bf q}^2)}{(t-m^{2}_{K})^{2}}
[\sigma_{\gamma \bar K^0\to K^{*-}\pi^+}(s_{1},t)],\nonumber\\
\frac{d\sigma_{\gamma p\to K^-\rho^+\Theta^+}}{dtds_1}
&=&\frac{g^{2}_{KN\Theta}}{32\pi^{2}sp^{2}_{i}}k\sqrt{s_{1}}
[-t+(m_N-m_{\Theta})^2]\frac{F({\bf q}^2)}{(t-m^{2}_{K})^{2}}
[\sigma_{\gamma \bar K^0\to\rho^+K^{-}}(s_{1},t)],
\end{eqnarray}
where $p_i$, $k$, and $s$ are similarly defined as in Eq.(\ref{threebody})
for proton-proton reactions with three-body final states. The form factor
$F({\bf q}^2)$ at $KN\Theta$ vertex is taken to have the same form
in Eq.(\ref{form}) with cutoff parameter $\Lambda=0.5$ GeV as used in
$\Theta^+$ production from meson-proton and proton-proton reactions.
Furthermore, we have introduced an overall monopole form factor for
two-body subprocesses $\gamma \bar K^0\to K^{*-}\pi^+$ and
$\gamma \bar K^0\to \rho^+ K^-$ with the same cutoff parameter
$\Lambda=0.75$ GeV like that used in charmed hadron production from
photon-proton reactions \cite{liu3}.

\begin{figure}[ht]
\includegraphics[width=3.5in,height=4.5in,angle=-90]{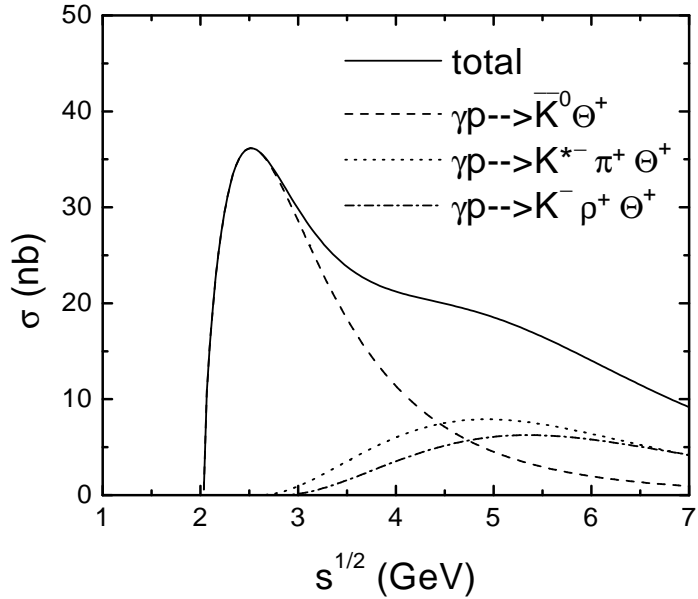}
\caption{Cross sections for $\Theta^+$ production from photon-proton
reactions as functions of center-of-mass energy: total (solid curve),
$\gamma p\to\bar K^0\Theta^+$ (dashed curve),
$\gamma p\to K^{*-}\pi^+\Theta^+$ (dotted curve), and
$\gamma p\to K^-\rho^+\Theta^+$ (dash-dotted curve).}
\label{cross3}
\end{figure}

In Fig.\ref{cross3}, we show the cross sections for $\Theta^+$
production from photon-proton reactions. It is seen that the cross
section with two-body final state, i.e., $\gamma p\to\bar K^0\Theta^+$
(dashed curve), dominates at low center-of-mass energies with peak
value about 36 nb, while those with three-body final states are more
important at high center-of-mass energies with maximum value
of 8 nb for $\gamma p\to K^{*-}\pi^+\Theta^+$ (dotted curve)
and 6 nb for $\gamma p\to K^-\rho^+\Theta^+$ (dash-dotted curve).
The total cross section including both two-body and three-body final
states is given by the solid curve.

\section{summary}
\label{summary}

The cross sections for the production of $\Theta^+$ baryon consisting
of $uudd\bar s$ quarks from meson-nucleon, proton-proton, and photon-proton
reactions are evaluated in a hadronic model that includes the
$KN\Theta$ interaction with coupling constant determined from the
width of $\Theta^+$. This model is based on a gauged SU(3) flavor
symmetric Lagrangian with the photon introduced as a $U_{\rm em}(1)$
gauged particle. Symmetry breaking effects are taken into account by
using empirical hadron masses and coupling constants. Form factors of
monopole type are introduced at interaction vertices to take into
account finite hadron sizes, and values of the cutoff parameters are
taken from fitting known cross sections of other reactions based on
similar hadronic models. It is found that for meson-nucleon reactions,
i.e., $\pi N\to\bar K\Theta^+$, $KN\to\pi\Theta$, and
$\rho N\to\bar K\Theta^+$, the one induced by kaon has the largest cross
section of about 1.5 mb and is almost an order-of-magnitude larger than
those for reactions induced by pion and rho meson. For proton-proton
reactions, the total cross section is about 20 $\mu$b and is dominated
by the reaction $pp\to\bar K^0 p\Theta^+$ with only about 25\% from
the reactions $pp\to\Sigma^+\Theta^+$ and $pp\to\pi^+\Lambda\Theta^+$.
In photon-proton reactions, the reaction $\gamma p\to\bar K^0\Theta^+$
with two-body final state is most important near threshold, and its
value is about 36 nb. At higher energy, the reactions
$\gamma p\to K^{*-}\pi^+\Theta^+$ and $\gamma p\to K^-\rho^+\Theta^+$
with three-body final states become important with comparable cross
sections of about 10 nb. Knowledge on these cross sections
is useful for studying $\Theta^+$ production not only in elementary
reactions involving hadrons and photons but also in relativistic heavy ion
collisions, where final hadronic effects on $\Theta^+$ production
and absorption need to be understood in order to infer its production
from the initial quark-gluon plasma.

\begin{acknowledgments}
This paper was based on work supported in part by the US National
Science Foundation under Grant No. PHY-0098805 and the Welch
Foundation under Grant No. A-1358.
\end{acknowledgments}

\end{document}